\documentclass[pr, reprint ,graphicx]{revtex4-1} 
\usepackage[fleqn]{amsmath}
\usepackage[]{amssymb}
\pdfoutput=1
\usepackage{color}
\usepackage[pdftex]{graphicx}
\usepackage{subfigure}
\begin{document} 
\title{Layered frustrated antiferromagnetic Heisenberg spin model: role of inplane frustration and interlayer coupling} 
\author{Md Mahfoozul Haque } 
\email{mahfoozulhaque@gmail.com} 
\affiliation{Department of Physics, Jamia Millia Islamia(Central  University), New Delhi 110025, India.}  
\author{M A H Ahsan}
\affiliation{Department of Physics, Jamia Millia Islamia(Central  University), New Delhi 110025, India.} 
\author{Jamal Berakdar}
\affiliation{Institute of Physics, Martin-Luther University, Halle-Wittenberg, 06120 Halle, Germany.}
\author{Dipan K Ghosh} 
\affiliation{Department of Physics, Indian Institute of Technology Bombay, Powai, Mumbai 400076, India.}
\date{\today}
\begin{abstract}
 We present an exact diagonalization study on layered $J_{1}-J_{2}$  antiferromagnetic Heisenberg spin model
to examine the role of frustration induced by inplane next-nearest neighbor coupling $J_{2}$, in presence of interlayer antiferromagnetic coupling $J_{\perp}$. A finite lattice of 24 spins in layered geometry of $(4\times 3)\oplus (4\times 3)$ is considered and the resulting Hamiltonian matrix diagonalized using Davidson iterative algorithm to obtain the ground and a few low-lying excited states. The lattice size ($24$ spins with Hilbert space dimensionality of $2704156$ in $S_{z}^{tot}=0$ subspace) has been kept relatively small because of the large number of runs required to sample $J_{1}-J_{2}-J_{\perp}$ parameter space. 
Quantities like spin-gap, Shannon entropy, spin-spin correlation(SSC), static spin structure-factor, magnetic specific-heat and magnetic susceptibility are calculated for various values of spin-spin coupling parameters. With increase in interlayer coupling, the system is driven to states with long range order and the interval of quantum paramagnetically disordered state,  sandwitched between N\'{e}el and collinear ordered states, narrows on the scale of inplane frustration parameter. 
\\ PACS: 75.10.Jm, 75.10.Kt, 75.40.-s 
\end{abstract}
\maketitle 
\section{Introduction}
The discovery of high temperature cuprate superconductors \cite{bednorz_zphysb_86_3}  triggered a renewed interest in quasi two-dimensional(2D) antiferromagnetic
Heisenberg spin model due to the belief that superconductivity is closely associated with the antiferromagnetic ordering in copper oxide planes in undoped parent compounds\cite{pwanderson_science_87_3, johnston_advphys_10_3}. 
In this regard, the antiferromagnetic $J_{1}-J_{2}$ quantum spin-$\frac{1}{2}$ model on 2D square lattice, with $J_{2}$ as frustration parameter, has been studied extensively\cite{bishop_prb_88, chandra_prb_88_3, dagotto_prl_89_3, richter_prb_93_3, Bishop_prb_98_3, capriotti_prl_00_3, capriotti_prl_01_3, singh_prl_03_3,rosner_prb_03_3, mahfooz_asl_15_3}.
The quantum fluctuation induced by frustration destroys the semi-classical N\'{e}el order, characterized by the magnetic wave-vector $(\pi,\pi )$. 
Beyond a critical value of $J_{2}$, the frustration drives the spin system to a new semi-classical order called the collinear order characterized by the magnetic wave-vector $(\pi,0)$ or $(0,\pi )$. 
At zero temperature, the quantum spin system is believed to have N\'{e}el order for $0\leq J_{2}/J_{1}<0.4$ and  collinear order for $0.6 < J_{2}/J_{1}\leq 1$. 
For intermediate values   $0.4 \leq J_{2}/J_{1}\leq 0.6$, the system is believed to exist in quantum  paramagnetically disordered state \cite{bishop_jpcm_08_3, bishop_prb_08_3}.
The boundary of the quantum paramagnetic$-$collinear order has  not been conclusively pinned down and there has been work to suggest it to be at $J_{2}/ J_{1}=0.65$ \cite{Schulz_epl_92_3,andrea_jpcm_96_3,Isaev_prb_09_3} or more.
\\ 
\indent
More recently, iron based superconductors have been found to have layered structure \cite{kamihara_jacs_08_3} and provide further motivation to study $J_{1}-J_{2}-J_{\perp}$ model \cite{melzi_prl_00_3,rosner_prl_02_3,abrahams_prl_08_3, sushkov_prb_11_3}. 
The role of $J_{\perp}$ on $J_{1}-J_{2}$ model has been examined through  several schemes like coupled-cluster, rotation-invariant Green's function \cite{richter_prl_06_3}, effective field theory\cite{nunes_condmat_10_3}, self consistent spin wave theory, series expansion and first order spin-wave theory\cite{sushkov_prb_11_3,stanek_prb_11_3,rojas_codmat_11_3}. 
For the antiferromagnetic 2D $J_{1}-J_{2}$ model in the intermediate regime of inplane next nearest neighbor coupling $0.4\leq J_{2}/J_{1}\leq 0.6$, it has been found that the quantum paramagnetic order narrows down \cite{richter_prl_06_3,sushkov_prb_11_3,nunes_condmat_10_3,stanek_prb_11_3,bishop_prb_88} with increase in interlayer coupling $J_{\perp}$. 
The critical value of $J_{\perp}$ at which the paramagnetic order disappears, has not been uniquely found and appears to depend strongly on the method of study employed \cite{richter_prl_06_3,sushkov_prb_11_3,nunes_condmat_10_3,stanek_prb_11_3,bishop_prb_88}.\\
\indent 
In the present work we use exact diagonalization to study, at zero as well as finite temperatures, the layered frustrated quantum spin-$\frac{1}{2}$ system with model Hamiltonian 
\begin{eqnarray}  
H &=& J_{1} \sum_{\left\langle i,j\right\rangle} {\bf S}_{i}\cdot {\bf S}_{j}+ 
J_{2}\sum_{\left\langle\left\langle i,j\right\rangle\right\rangle}{\bf S}_{i}\cdot {\bf S}_{j} + J_{\perp}\sum_{\left\langle\left\langle\left\langle i,j\right\rangle\right\rangle\right\rangle}{\bf S}_{i}\cdot {\bf S}_{j}\nonumber \\
\label{hamiltonian}
\end{eqnarray}
where $\left\langle i,j\right\rangle $ and $\left\langle\left\langle i,j\right\rangle\right\rangle $ are inplane nearest and next nearest neighbors respectively and  $\left\langle\left\langle\left\langle i,j\right\rangle\right\rangle\right\rangle $ represents the interplane  nearest neighbors; the couplings $J_{1},\ J_{2},\ J_{\perp}$ being the isotropic exchange integrals for the respective neighbors which have been taken to be positive.\\
\indent 
 In order to validate our exact diagonalization code, we carried out diagonalization on $16$ site spin-$\frac{1}{2}$ chain with nearest neighbor coupling $J_{1}$ only and found the ground state energy per spin to be $-0.446393523 J_{1}$, in agreement with Bethe Ansatz results. 
\begin{table}[h!] 
\centering
\setlength{\tabcolsep}{5pt}
\begin{tabular}{|l|ccc|}
\hline
$J_{2}$ & $E_{0}(S_{tot})$ & $E_{1}(S_{tot})$ & $E_{2}(S_{tot})$ \\
\hline
$0.85 $ & $-1.088789(0)$ & $-1.008873(0)$ & $-0.995889(1)$ \\
$0.90 $ & $-1.076600(0)$ & $-1.011288(0)$ & $-0.979593(1)$ \\
$0.95 $ & $-1.065978(0)$ & $-1.014114(0)$ & $-0.963962(1)$ \\
$1.00 $ & $-1.057240(0)$ & $-1.017374(0)$ & $-0.953479(0)$ \\
$1.05 $ & $-1.050794(0)$ & $-1.021119(0)$ & $-0.954172(0)$ \\
$1.10 $ & $-1.047189(0)$ & $-1.025481(0)$ & $-0.970216(0)$ \\
$1.15 $ & $-1.047183(0)$ & $-1.030808(0)$ & $-0.989355(0)$ \\
$1.20 $ & $-1.051792(0)$ & $-1.038178(0)$ & $-1.008256(0)$ \\
\hline
\end{tabular}
\caption{The ground state and first two excited-state energies  $(E_{0}, E_{1}, E_{2})$ per spin, with total spin of the respective states given in parenthesis, on a $4\times 4$ two-dimensional lattice for spin-$\frac{1}{2}$ Heisenberg antiferromagnet for several values of next nearest-neighbor exchange coupling $J_{2}$. 
Here $J_{1}$ is taken to be 2, as in reference \cite{dagotto_prb_89_3}.}
\label{4x4_latiice_results}
\end{table}
To further test the validity of the code on 2D lattice, we  diagonalized the spin-$\frac{1}{2}$ Hamiltonian for $4\times 4$ lattice with nearest and next nearest neighbor couplings $J_{1}$ and $J_{2}$. 
The ground and first two excited state energies per spin along with total spin of the respective states are given in Table \ref{4x4_latiice_results} and is in agreement with the results of \cite{dagotto_prb_89_3}.
The same code has earlier been used to analyze the quasi-1D characteristics of Sr$_{2}$CU(PO$_{4}$)$_{2}$ and Ba$_{2}$CU(PO$_{4}$)$_{2}$ \cite{mahfooz_jmmm_16_3} and the calculated values were found in good agreement with experimental results. 
\\
\indent 
This paper is organized as follows. 
In section II, we present the physical quantities that are to be calculated in subsequent sections. Section III describes the $24-$site finite lattice for our model calculation.
In section IV, we present our numerical results on the finite lattice considered for $J_{1}-J_{2}-J_{\perp}$ quantum spin-$\frac{1}{2}$ model.  
Results are summarized and discussed in section V.
\section{Physical quantities calculated}
In order to obtain finite temperature quantities, we require to calculate the canonical partition function 
\begin{eqnarray}
 \label{partition}
 Z_{c}(T,N) =Tr\left( e^{-\beta H}\right) =  \sum_{\alpha=1}^{d} e^{-\beta E_{\alpha}}
\end{eqnarray}
where \ $d$ is the dimensionality of the many-body Hilbert space. 
For the antiferromagnetic system, the order parameter, staggered magnetization, is  defined as \cite{henley_prl_62}
\begin{eqnarray}
{{\bf m}_{s}}\left({\bf Q}\right) = \sum_{i}e^{i {\bf Q}\cdot {\bf r}_{i}}\ {\bf S}_{i} \nonumber
\end{eqnarray}
where {\bf Q} is the magnetic wave-vector and the sum on the right hand side is over all the lattice sites. 
In the thermodynamic limit, $\langle{\bf m}_{s}\rangle$ is non-zero in the ordered phase and zero in the disordered phase where the angular brackets $\langle \cdots \rangle $ denote the ensemble averaging. 
For a finite system, however,  $\langle{\bf m}_{s}\rangle$ vanishes for all phases due to rotational symmetry in spin space and it becomes more appropriate to consider the square of the order parameter, the static spin structure-factor defined as 
\begin{eqnarray}
S({\bf Q}, T)&=&\frac{1}{N}\langle{\bf m}_{s}\cdot{\bf m}_{s}\rangle 
=\frac{1}{N}\sum_{i,j}e^{{\bf Q}\cdot({\bf r}_{i}-{\bf r}_{j})}\left\langle 
{\bf S}_{i}\cdot {\bf S}_{j}\right\rangle \\
&=& \frac{1}{N}\sum_{i,j } e^{i {\bf Q}\cdot({\bf r}_{i} - {\bf r}_{j})} 
\sum_{\alpha=1}^{d}\frac{e^{-\beta E_{\alpha}}}{Z_{c}\left(T,N\right)}\left\langle 
E_{\alpha}\left|{\bf S_{i}} \cdot {\bf S_{j}}\right|E_{\alpha}\right\rangle \nonumber
\label{structurefactor}
\end{eqnarray}
where the sums $i,j$ run over all the $N$ lattice points. 
The quantity $\left\langle E_{\alpha}\left|{\bf S}_{i}\cdot {\bf S}_{j}\right|E_{\alpha} \right\rangle$ represents the SSC between $i^{th}$ and $j^{th}$ spins for the $\alpha^{th}$ energy eigenstate. 
At zero temperature, the SSC becomes 
\begin{eqnarray} 
&& \left\langle{\bf S}_{o} \cdot {\bf S}_{r} \right\rangle_{0} = \left\langle E_{0} \left| S_{0}^{z}S_{r}^{z} +\frac{1}{2}\left ( S_{o}^{+}S_{r}^{-}+S_{o}^{-}S_{r}^{+}\right )\right| E_{0}\right\rangle 
\label{structure_factor_0}
\end{eqnarray} 
where $\left| E_{0}\right\rangle$ is the many-body ground state. The spins ${\bf S}_{o}$ and ${\bf S}_{r}$ are the reference spin and its $r^{th}$ neighbor respectively.\\
\indent 
Another thermodynamic quantity of interest is the magnetic specific heat given as 
\begin{eqnarray}
\frac{C}{k_{B}} &=& \beta^{2} \frac{\partial^{2}\ln Z_{c}\left(T,N\right)}{\partial \beta^{2}} \qquad \mbox{where} \quad \beta= \frac{1}{k_{B}T} \nonumber \\
&=&\beta^{2}\sum_{\alpha}\{2S_{tot}(E_{\alpha})+1\}E_{\alpha}^{2}\frac{e^{-\beta E_{\alpha}}}{Z_{c}}  \nonumber \\ 
&& \qquad \times \left\{1 - \{2S_{tot}(E_{\alpha})+1\} \frac{e^{-\beta E_{\alpha}}}{Z_{c}} \right\}
\label{specificheat} 
\end{eqnarray}
where $S_{tot}(E_{\alpha})$ is total spin of $\alpha^{th}$ eigenstate corresponding to eigenvalue $E_{\alpha}$ \\
\indent
The $z$-component of the magnetic susceptibility for an isotropic system is given as \cite{haraldsen_prb_05_3} 
\begin{eqnarray}
\label{chi}
\chi^{zz} &=& \frac{\beta}{Z_{c}\left(T,N\right)}\sum_{\alpha=1}^{d}\left({\cal M}_{z}^{2} \right)_{\alpha}e^{-\beta E_{\alpha}} \quad \mbox{with } {\cal M}_{z} = \frac{g \mu_{B}}{\hbar} S^{z}_{tot}\nonumber \\
 &=& \frac{1}{3}\left(g \mu_{B} \right)^{2} \frac{\beta}{Z_{c}\left(T,N\right)}
\sum_{E_{\alpha}}S_{tot}(E_{\alpha})\left\{2 S_{tot}(E_{\alpha})+1 \right\} \nonumber \\
 && \qquad \qquad \qquad \quad\times \left\{S_{tot}(E_{\alpha})+1 \right\}e^{-\beta E_{\alpha}}
\end{eqnarray}
where $S_{tot}$ is the total spin of the eigenstate, $\mu_{B} $ the Bohr magneton and $g$ the electron Lande g factor. 
\section{The finite lattice for model calculation}
\subsection{The Finite Lattice}
\begin{figure}[h!]
\centering
\includegraphics[width=0.45\textwidth]{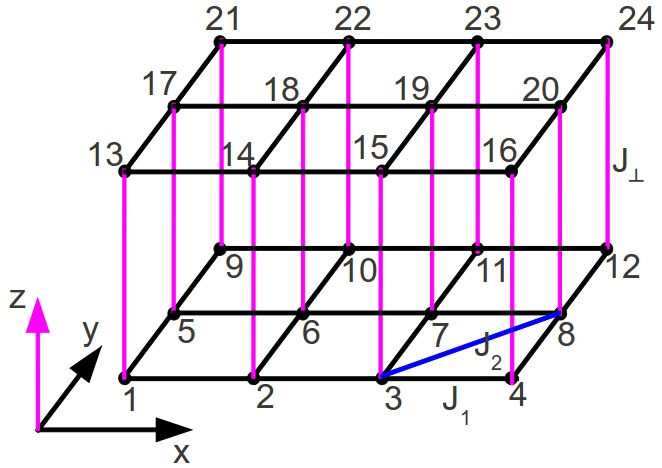}
\caption{(Color online) The 24 spin two-layered cubic  lattice in $(4\times 3)\oplus(4\times 3)$ geometry. 
The exchange couplings $J_{1}$, $J_{2}$ are the nearest and the next nearest neighbor inplane couplings respectively and $J_{\perp}$ the nearest neighbor interlayer coupling. 
Periodic boundary condition is used along $x-$direction but along $y-$direction, open boundary condition has been used. 
The value of $J_{\perp}$ in the present calculation is taken double its value in a multi-layered lattice (see text).}
\label{24spn_lattice}
\end{figure}

We consider a two-layered lattice containing 24 spins with 12 spins in each two-dimensional layer in ($4\times 3$) geometry shown in Fig.(\ref{24spn_lattice}). 
It will, henceforth, be referred to as $(4\times 3)\oplus(4\times 3)$ lattice with $J_{1}$, $J_{2}$ being the inplane nearest and next nearest neighbor couplings respectively and $J_{\perp}$ the interlayer coupling. 
Each layer is described by a $J_{1}-J_{2} $ frustrated antiferromagnetic Heisenberg(FAFH) spin model. 
In the two-layered lattice considered here, there are four spins in the $x$-direction and three spins in the $y$-direction. 
We, therefore, use periodic boundary condition (PBC) along $x$-direction and, in order to avoid geometrical frustration, open boundary condition (OBC)
along $y-$direction . 
Since there are only two layers along the $z-$direction, use of PBC leads to double counting of the interlayer coupling and hence, in our calculation, $J_{\perp}$ is taken twice its value used in a multi-layered system studied, say, in reference\cite{richter_prl_06_3} where several layers of $J_{1}-J_{2}$ model has been examined  using coupled-cluster method.  
The Hilbert space dimensionality for a spin-$\frac{1}{2}$ system on a $24$-site lattice is $2^{24} = 16777216\approx10^{7}$. 
As discussed in the next subsection, it suffices to work in ${\bf S}^{z}_{tot}=0$ subspace, which has projections from all the total spin states ${\bf S}_{tot}$ and  leads dimensionality of the Hamiltonian matrix reduce to $2704156$, which we diagonalize iteratively to obtain the ground and a few low-lying excited states using Davidson algorithm for large sparse matrices. 
\subsection{The Ising basis}
It is readily seen that the Hamiltonian in Eq(\ref{hamiltonian}) commutes with the square of the total spin ${\bf S}^{2}_{tot}$ and one of its components ${\bf S}^{z}_{tot}$
\begin{eqnarray*}
\left[ {\bf S}^{2}_{tot},  H\right]=0, \ \ \ \left[{\bf S}^{z}_{tot},
 H \right]=0, \ \ \mbox{   with  } \ \ {\bf S}_{tot}= \sum_{i=1}^{N}
{\bf S}_{i}
\label{constantofmotion}
\end{eqnarray*}
and hence are constants of motion with eigenvalues $S_{tot}\left(S_{tot}+1\right)$ and $M_{s}^{tot}$, respectively. 
For a system with even number of spins, it suffices to perform the diagonalization of the Hamiltonian (\ref{hamiltonian}) in $M_{s}^{tot}=0$ subspace which has projections from all the total spin subspaces $S^{tot}=\frac{N}{2}, \cdots , 1, 0$. 
For an $N-$spin system, a many-body Ising basis state in a given total $S^{z}_{tot}$ subspace is given by 
\begin{eqnarray}
\left| N, M_{s}^{tot}; k\right\rangle =\left|m_{1},m_{2},\cdots
 m_{N};\sum_{i=1}^{N}m_{i}=M_{s}^{tot}\right\rangle
\label{isingbasis}
\end{eqnarray}
where $k$ runs over the dimensionality of the $N$-spin Hilbert space in the given $S_{tot}^{z}$ subspace.
\section{Numerical Results}
In the numerical results presented here, all the exchange couplings and energies have been measured in units of the inplane nearest-neighbor coupling $J_{1}$. \\
\subsection{The spin-gap}
\begin{figure}[h!]
\centering
\includegraphics[width=0.5 \textwidth]{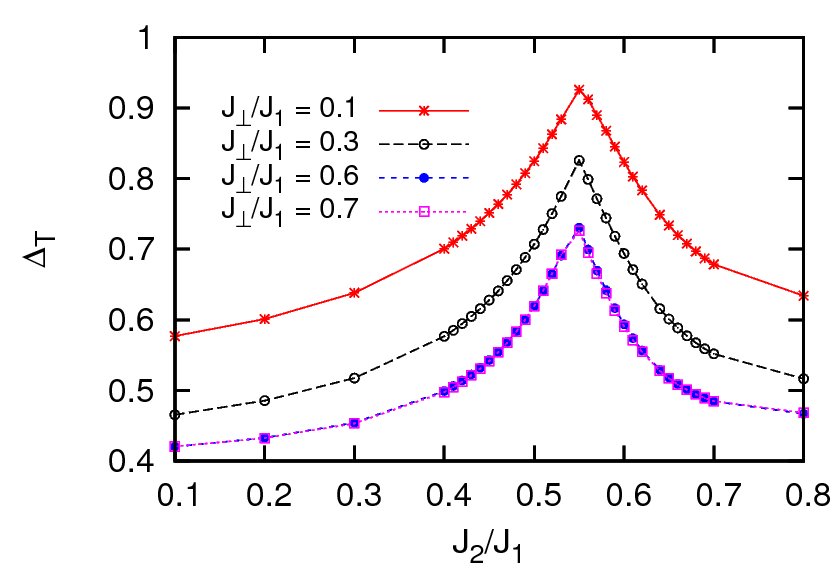}
\caption{Spin-gap $\Delta_{T}$ {\it vs} inplane next nearest-neighbor coupling $J_{2}/J_{1}$, at different values of interlayer coupling $J_{\perp}/ J_{1}$ for the two-layered $\left(4\times 3\right)\oplus\left(4\times 3\right)$ lattice. The values of $J_{\perp}$ taken in the above graph is twice the values used in reference\cite{richter_prl_06_3} which considers a multi-layered lattice. 
The spin-gap $\Delta_{T}$ attains a maximum with peak around $J_{2}/J_{1} \approx 0.55$, for the four values of interlayer coupling $J_{\perp}/J_{1}$ considered here.}
\label{24spn_spin_gap} 
\end{figure}

The spin-gap is defined as the difference of energies of the lowest triplet state and the singlet ground state \cite{richter_epjb_10_3}
\begin{eqnarray}
\Delta_{T}&=&E_{0}\left(S_{tot}=1\right)-E_{0}\left(S_{tot}=0\right) 
\label{spingapdef}
\end{eqnarray}
 where $E_{0}\left(S_{tot}=1\right) $ and $E_{0}\left(S_{tot}=0\right)$ are the lowest eigenenergies in the subspace  $S_{tot}=1$ and $E_{0}\left(S_{tot}=0\right)$, repspectively.\\
\indent
It has been reported that the spin-gap has non-zero values in quantum paramagnetic regime of a 2D $J_{1}-J_{2}$ model\cite{richter_epjb_14_3}. In Fig.(\ref{24spn_spin_gap}), we present spin-gap $\Delta_{T}$ {\it vs} inplane frustration parameter in the interval $0\leq J_{2}/J_{1}\leq 0.8$ for different values of interlayer coupling $J_{\perp}/J_{1}$. 
We observe that, at a given value of $J_{\perp}/J_{1}$, the spin-gap ${\Delta_T}$ increases slowly, attains a maximum, with (a diverging) peak at $J_{2}/J_{1} \approx 0.55$ and then decreases with the inplane frustration parameter,  indicating the existance of a gapful paramagnetically disordered state \cite{qmag_ssachdev_3} in the interval $0.4< J_{2}/J_{1}< 0.7$. 
For small value of $J_{\perp}/J_{1}=0.1$, we get ${\Delta_T}/N \approx 0.04$  at $J_{2}/J_{1}=0.55$ in agreement with the vlaue of ${\Delta_T}(\infty)$ reported in \cite{gong_prl_14_3} for an open rectangular cylinderical lattice.
As the interlayer coupling $J_{\perp}/J_{1}$ is increased, the $\Delta_{T}$ {\it vs} $J_{2}/J_{1}$ curve shifts downward i.e. the value of spin-gap decreases with increase in interlayer coupling. 
The peak of the plots becomes narrower with increase in $J_{\perp}/J_{1}$, over the frustration parameter interval $0.4 < J_{2}/J_{1} < 0.7$. This indicates that the region of gapful paramagnetically  disordered regime shrinks on $J_{2}/J_{1}$ scale with increase in $J_{\perp}/J_{1}$. 
We further observe that the $\Delta_{T}$ {\it vs} $J_{2}/J_{1}$ plots for $J_{\perp}/J_{1}=0.7 $ and $J_{\perp}/J_{1}=0.6 $ coincide {\it i.e.} an increase in the interlayer coupling beyond $J_{\perp}/J_{1}=0.6 $  does not affect the spin-gap. 
It is to be stated here that since our exact diagonalization calculation considers only two layers of $J_{1}-J_{2}$ model and double count the interlayer coupling, the value of $J_{\perp}/J_{1}$ in our case is indeed {\em twice} the value in a multi-layered lattice considered, for example, in reference\cite{richter_prl_06_3}. 
It has been reported in reference \cite{richter_prl_06_3} that the paramagnetically disordered state vanishes at $J_{\perp}=0.3J_{1}$ and the system goes directly from semi-classical N\'{e}el order to semi-classical collinear order.
\subsection{Spin-spin correlation: Zero temperature}
\begin{figure}[h!]
\centering
\subfigure[The inplane first-neighbor ($5th$ and $6th$ SSC {\it vs} $J_{2}/J_{1}$ for different values of interlayer coupling $J_{\perp}/J_{1}$. As the $J_{\perp}/J_{1}$ is increased, the SSC exhibit significant decrease in inplane frustration parameter interval $0.4\le J_{2}/J_{1}\le 0.7$, indicating the diminishing of the paramagnetically disordered state.] 
{\includegraphics[width=0.45\textwidth]{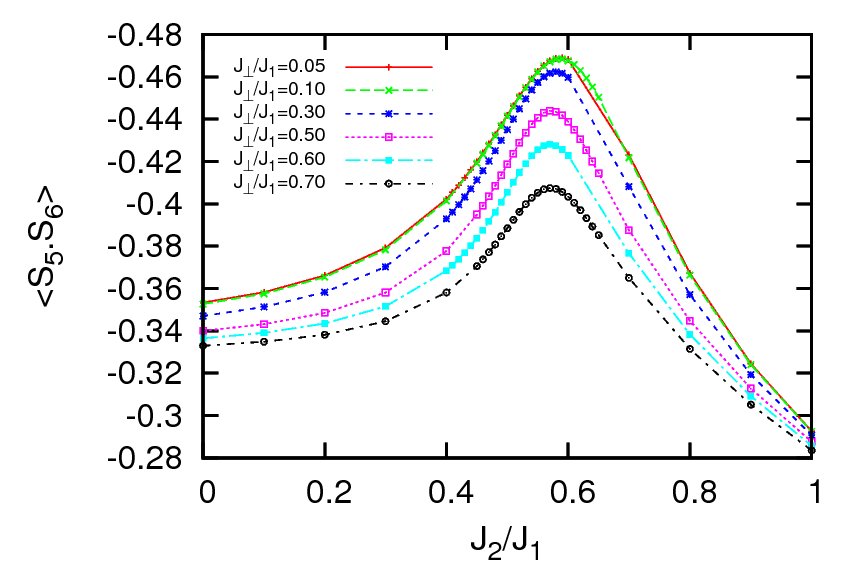}} \qquad
\subfigure[The inter-plane (along $z$-direction) first neighbor SSC between $5th$ and $17th$ spins (refer to Fig.\ref{24spn_lattice}) {\it vs} $J_{2}/J_{1}$ for various values of $J_{\perp}/J_{1}$. As the interlayer coupling is increased significantly to, say $J_{\perp}/J_{1}=0.7$, the inter-plane SSC function attains a (negative) peak in the interval $0.4\le J_{2}/J_{1}\le 0.7$ . ]
{\includegraphics[width=0.45\textwidth]{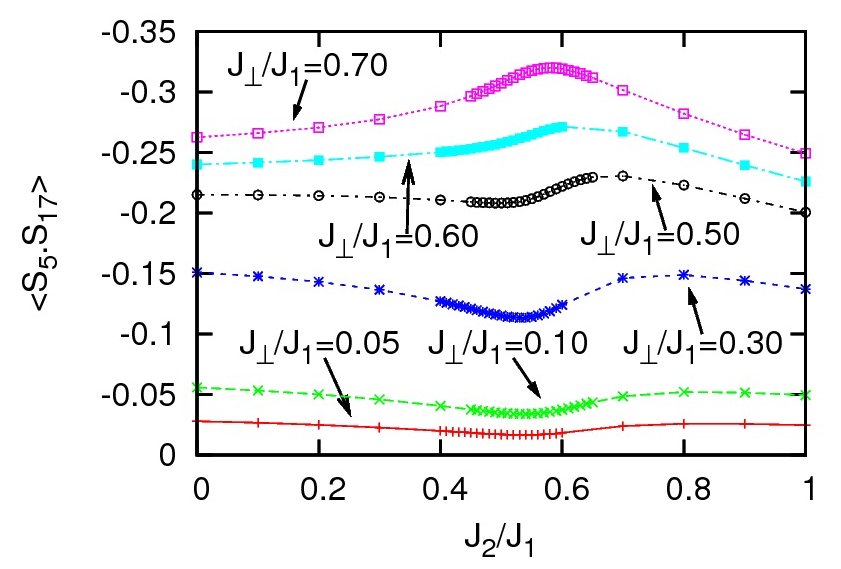}}
\caption{Inplane and inter-plane first neighbor SSC function 
{\it vs} $J_{2}/J_{1}$ for different values of interlayer 
coupling $J_{\perp}/J_{1}$. }
\label{24spn_spn_cor_1st_nbr}
\end{figure}

In order to show the consistency of our results with the known results regarding various orders of 2D $J_{1}-J_{2}$ lattice on $J_{2}/J_{1}$ scale, we consider the static SSC for the ground state and its variation with $J_{2}/J_{1}$ as well as $J_{\perp}/J_{1}$. 
\begin{figure}[h!]
\centering
\subfigure[ Inplane SSC {\it vs} frustration parameter $J_{2}/J_{1}$ with interlayer coupling $J_{\perp}/J_{1}=0.3$. Spin 5 has been 
chosen as the reference spin and $n = 6, 10, 7, 11$ correspond to first, second, third and fourth neighbors respectively.]
{\includegraphics[width=0.5\textwidth]{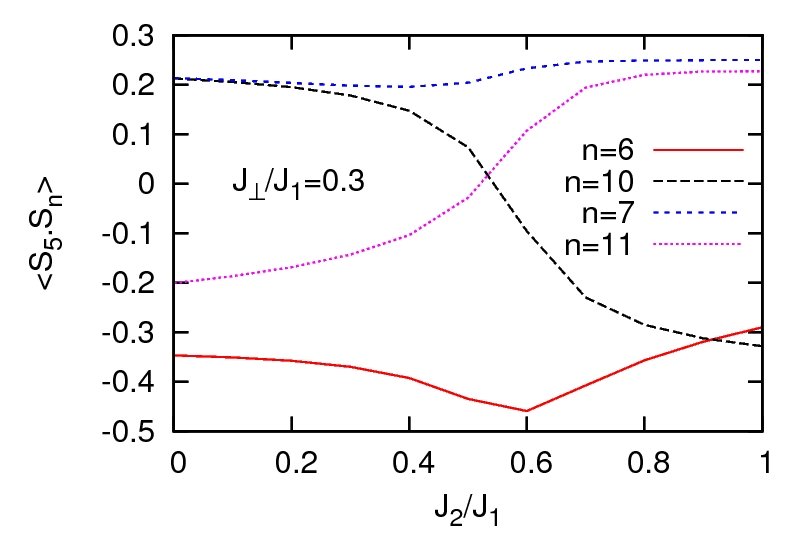}} 
\subfigure[Arrangement of spins on a $4\times 3$ lattice in a layer with PBC along $x$-direction and OBC along $y$-direction for (i) semiclassical N\'{e}el order corresponding to small values of $J_{2}/J_{1}$, and (ii) semiclassical collinear order corresponding to large values of $J_{2}/J_{1}$.]
{\includegraphics[width=0.5\textwidth]{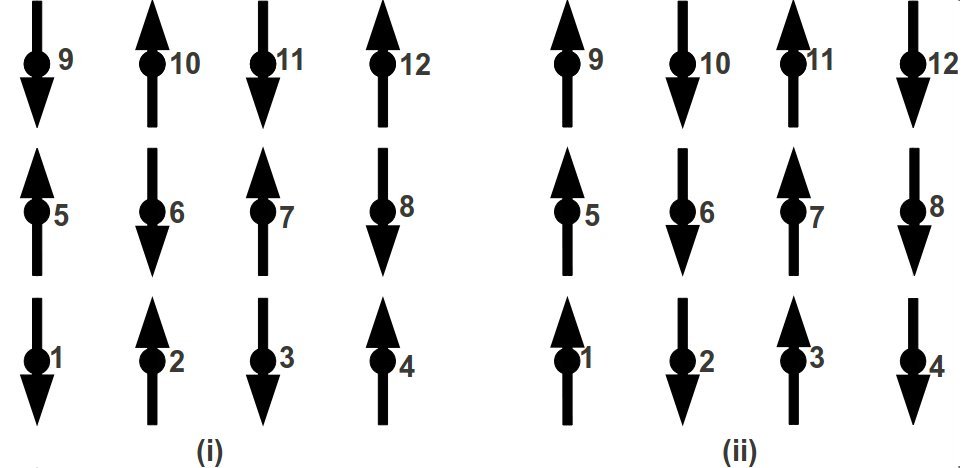}} 
\caption{Inplane SSC {\it vs} frustration parameter $J_{2}/J_{1}$ and spin arrangements for semiclassical N\'{e}el and collinear orders.}
\label{24spncorr_zpt3}
\end{figure}
In Fig.(\ref{24spncorr_zpt3}a), we present the variation  of the inplane SSC with $J_{2}/J_{1}$ for the first(n=6), second(n=10), third(n=7) and fourth(n=11) neighbors (Fig.(\ref{24spncorr_zpt3}b)). 
The inplane first(n=6) and third(n=7) neighbors along $x$-axis along which PBC has been used, the SSC takes negative values (anti-parallel spins) and positive values (parallel spins), respectively, for all values of $J_{2}/J_{1}$, {\it i.e.} spin $6$ and spin $7$ remain antiparallel and parallel, respectively with respect to the reference spin 5, irrespective of whether the system is in the semiclassical N\'{e}el or the semiclassical collinear state. 
However, the second(n=10) and the fourth (n=11) neighbor SSC goes from positive to negative and negative to positive values, respectively crossing each other at $J_{2}/J_{1}\approx 0.55$, as  $J_{2}/J_{1}\in [0,1]$ is increased. 
That is to say, for values of $J_{2}/J_{1}<0.4 $, a spin system on 2D square lattice exists in semiclassical N\'{e}el ordered state whereas for values of $J_{2}/J_{1}>0.7$, the system acquires semiclassical collinear order. 
These results are consistent with known results for 2D $J_{1}-J_{2}$ FAFHM spin model \cite{reuther_prb_11_3}.\\
\indent
Fig.(\ref{24spn_spn_cor_1st_nbr}a) shows the variation of inplane nearest neighbor correlation between the $5th$ and $6th$ spins with $J_{2}/J_{1}$ for various values of $J_{\perp}/J_{1}$. 
The inplane nearest-neighbor SSC attains a 
(negative) maximum between $0.4 \le J_{2}/J_{1} \le 0.7 $ indicating that the short range correlation (disorder) is dominant in the said interval of frustration parameter $J_{2}/J_{1}$. It is also seen that as the interlayer coupling $J_{\perp}/J_{1}$ is increased, the height of (negative) peak decreases, {\it i.e.} short range correlation decreases with increase in $J_{\perp}/J_{1}$ in the interval $0.4 \le J_{2}/J_{1}\le 0.7$.\\
\indent 
Fig.(\ref{24spn_spn_cor_1st_nbr}b) presents the interplane first-neighbor correlation 
between $5th$ and $17th$ spins 
{\it vs} inplane frustration 
parameter $J_{2}/J_{1}$ for several values of interlayer coupling. The SSC $\langle S_{5}\cdot S_{17} \rangle$ increases with increase in $J_{\perp}/J_{1}$ for all values of inplane frustration parameter in the interval $0\le J_{2}/J_{1} \le 1$. For small values of $J_{\perp}/J_{1}$, the 
inter-plane SSC changes very little with $J_{2}/J_{1}$ indicating that the physics is essentially the same as that of 2D $J_{1}-J_{2}$ model. However, when the interlayer coupling is made significant {\it i.e.} $J_{\perp}/J_{1}>0.10$, the inter-plane first neighbor SSC attains a peak between $0.4 \le J_{2}/J_{1}\le 0.7$ with accompanying decrease in inplane first neighbor SSC, indicating the shrinking of quantum disorder in the said interval of inplane frustration parameter, as seen in Fig.(\ref{24spn_spn_cor_1st_nbr}a).
\subsection{Zero temperature static spin structure-factor}
\begin{figure}[h!]
\centering
\includegraphics[width=0.5\textwidth]{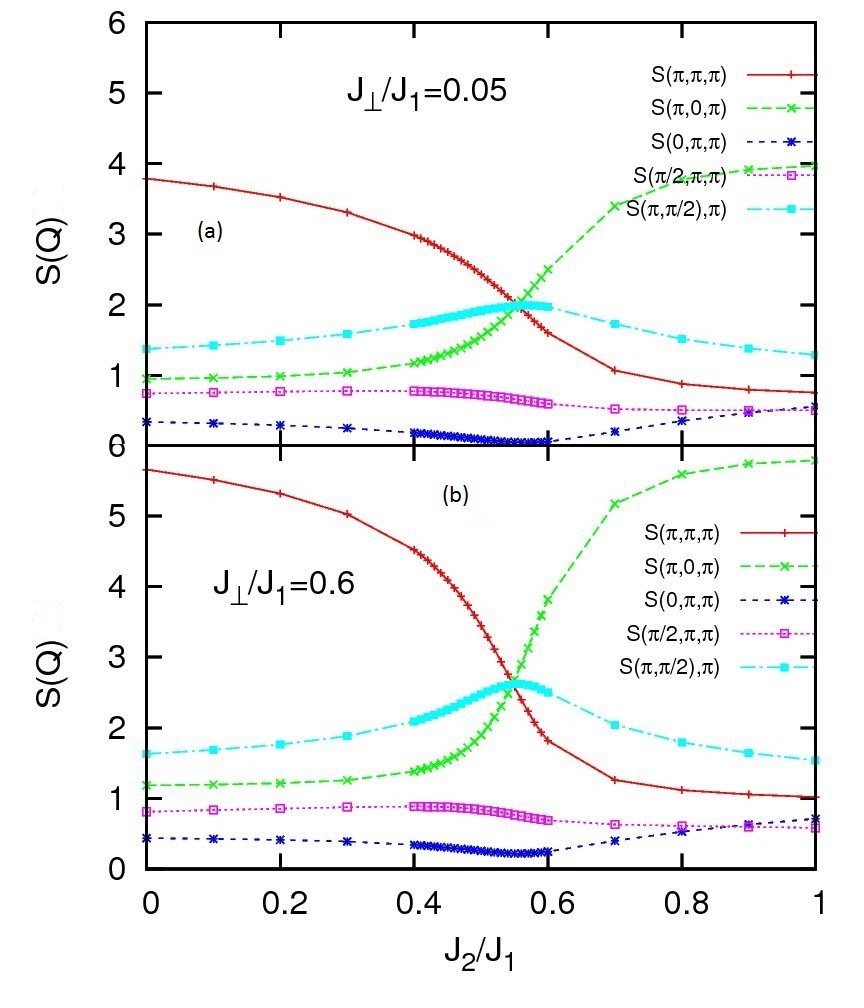} 
\caption{Zero temperature static spin structure-factor $S\left(Q_{x},Q_{y},\pi\right)$ {\it vs} $J_{2}/J_{1}$ for the layered $\left(4\times 3\right)\oplus\left(4\times 3 \right)$ lattice at two different values of $J_{\perp}/J_{1}=0.05,\ 0.6$. The quantity $S\left(\pi,\pi,\pi\right)$ 
corresponding to the N\'{e}el order is dominant at small values of 
$J_{2}/J_{1}$, whereas for large values of $J_{2}/J_{1}$, $S\left(\pi,0,\pi\right)$ corresponding to the inplane collinear order is dominant. For intermediate values $0.4 \le J_{2}/J_{1}\le 0.7$, several of the competing magnetic orders coexist leading to quantum paramagnetically  disordered state.} 
\label{24spn_sf_zpt6&pt05}
\end{figure}
Another quantity we examine is the zero-temperature static spin structure-factor $S({\bf Q})$ defined in equation(\ref{structure_factor_0}). 
In Fig.(\ref{24spn_sf_zpt6&pt05}), we present $S({\mathbf Q})$ {\it vs} $J_{2}/J_{1}$ for two different values of 
$J_{\perp}/J_{1}=0.05,\ 0.6$, for various values of the 
magnetic wave-vector {\bf Q}. For small values of $J_{2}/J_{1}$, the static 
spin structure-factor $S(\pi,\pi,\pi)$ corresponding to the N\'{e}el order 
dominates whereas for large values of $J_{2}/J_{1}$, the spin-structure factor 
$S\left(\pi ,0,\pi\right)$ corresponding to the collinear order dominates. It is to be noted that in our calculation 
$S\left(\pi,0,\pi\right)\neq\left(0,\pi,\pi\right)$, because PBC has been used along $x$-direction only, whereas OBC has been used along $y$-direction. For intermediate values of frustration parameter in the interval $0.4\le J_{2}/J_{1}
\le 0.7$, the spin structure factor $S\left({\mathbf Q}\right)$ corresponding 
to  several magnetic wave-vectors ${\mathbf Q}$ have comparable values, implying that several of the competing magnetic orders coexist in the said interval leading to quantum paramagnetically disordered state.
As the interlayer coupling is increased to a large 
value, say $J_{\perp}/J_{1}=0.6J_{1}$, the spin structure factor 
$S(\pi,\pi,\pi)$ corresponding to N\'{e}el ordering takes markedly large values for $J_{2}/J_{1}\le 0.4$ whereas for $J_{2}/J_{1}\ge 0.7$, it is the spin-structure factor $S(\pi,0,\pi)$ corresponding to collinear order that takes markedly large values, as seen in Fig.(\ref{24spn_sf_zpt6&pt05}b). For intermediate values of inplane frustration parameter $0.4\leq J_{2}/J_{1}\leq 0.7$, the spin structure factors $S\left(\pi,\pi,\pi\right)$ and $S\left(\pi ,0,\pi\right)$ vary very sharply and the region of quantum paramagnetically disordered state becomes narrower on $J_{2}/J_{1} $ scale. This can be seen as the paramagnetically disordered state tending to disappear with increase in interlayer coupling.
\subsection{Shannon Entropy}
At zero temperature, the quantum mechanical ground state of the Frustrated Antiferromaggnetic spin-$\frac{1}{2}$ system described by the Hamiltonian
in equation (\ref{hamiltonian}) can be viewed as a 
statistical mixture of Ising basis states, equation(\ref{isingbasis}).
Accordingly, the Shannon entropy (SE), widely used in information theory, can 
be defined  \cite{shannon_bell_sys_tech_48_2} for a quantum mechanical state $\left|\Psi\right\rangle=\sum_{i} c_{i}\left|N,M_{s}^{tot};i\right\rangle $ as 
\begin{eqnarray}
\frac{S}{N k_{B}}= -\sum_{i}p_{i}\ln p_{i}
\end{eqnarray}
where $N$ is the number of spins and 
\begin{eqnarray}
p_{i} &=& \left|c_{i}\right|^{2} \ \ \ \ \ \ \mbox{ with }\ \ \   
\sum_{i} \left|c_{i}\right|^{2}~=~1 , 
\end{eqnarray}
is the probability of the $ith$ Ising basis state in the variational 
wavefunction $\left|\Psi\right\rangle $ obtained through exact diagonaliztion of the Hamiltonian matrix. 
Since there are $2^{N}$ basis states for $N$ spin$-\frac{1}{2}$ particles, $p_{i}=\left(1/2^{N}\right)$ for maximally disordered state  and the maximum value of $S/(N k_{B})$ is $\ln 2 =0.693$.
\begin{figure}[h!]
\includegraphics[width=0.45\textwidth]{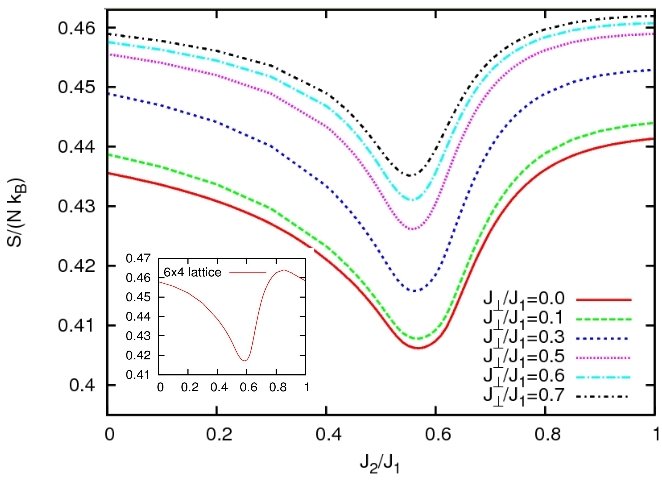}
\caption{SE of $(4\times 3)\oplus (4\times 3)$ lattice {\it vs} $J_{2}/J_{1}$  for different values of inter layer couplings $J_{\perp}/J_{1}$. 
At $J_{\perp}/J_{1}=0$, two layers becomes separated therefore we present the, SE {\it vs} $J_{2}/J_{1}$, for a single layer $(4\times 3)$ lattice. 
The inset image displays the variation of SE with $J_{\perp}/J_{1}$ for single layer $\left( 6\times 4 \right)$ lattice, $J_{1}-J_{2}$ model.}
\label{shannon_entropy}
\end{figure}
In Fig.(\ref{shannon_entropy}) we present the variation of Shanon Entropy(SE)
 with inplane frustration parameter $J_{2}/J_{1}$ for different values of 
interlayer couplin $J_{\perp}/J_{1}$. With the interlayer 
coupling taken to zero, the system becomes a set of two 
independent layers of $(4\times 3)$ lattice with PBC
 along $x-$direction and OBC along $y-$direction. 
The SE {\it vs} $J_{2}/J_{1}$ plot for a $(4\times 3)$ lattice has been shown by red solid line and labeled as $J_{\perp}/J_{1}=0.0$ in Fig.(\ref{shannon_entropy}). For a better comparison with a more realistic two-dimensional lattice, we present in inset, SE {\it vs} $J_{2}/J_{1}$ plot for a $24-$spin $\left(6\times 4\right)$ lattice in which PBC has been imposed both along $x-$ and $y-$ directions. 
In the parameter intervals $0\leq J_{2}/J_{1}<0.4$ and $0.7 < J_{2}/J_{1} \leq 1$, we obtain large values of Shanon Entropy(SE) implying that the system fluctuates around the semiclassical N\'{e}el state and the semiclassical collinear state, respectively. The shanon entropy is minimum at a value of inplane frustration parameter $J_{2}/J_{1}$ a little less than 0.6, implying that some kind of order-by-quantum-disorder\cite{yildirim_tjp_99_3, ross_prl_112_057201} takes 
place in spin configuration space in the parameter interval $0.4<J_{2}/J_{1} <0.7$. 
As we increase the $J_{\perp}/J_{1}$, values of SE increases in the regime of small and large values of $J_{2}/J_{1}$, whereas the depth of the the plot in the intermediate regime $0.4<J_{2}/J_{1} <0.7$ corresponding to the disordered state takes further dip. This indicates the narrowing of intermediate quantum mechanically disordered regime on $J_{2}/J_{1}$ scale, with increase in interlayer coupling \cite{richter_prl_06_3,sushkov_prb_11_3,nunes_condmat_10_3,stanek_prb_11_3}.  
\subsection{Specific heat and magnetic susceptibility}
In order to calculate a thermodynamic quantity, we require, in principle, the ground state and all the excited states. This may be computationally  prohibitive for 
 most of the systems of interest. However, quantum fluctuation arising due to frustration is important only at low temperatures where only a few low-lying excited states are relevant to any thermodynamic quantity. 
 We, in the present calculation, take only the lowest six eigenstates, to examine the effect of $J_{\perp}$ on specific heat and magnetic susceptibility for the $(4\times 3)\oplus(4\times 3)$ lattice.
We find that for the Hamiltonian in equation(\ref{hamiltonian}), the lowest six eigenstates belong to $S_{tot}= 0,\ 1$ and $2$ subspace in the parameter regimes $0 \le J_{2}/J_{1} \le 0.4 $ and $0.7 \le J_{2}/J_{1} \le 1$ whereas in the intermediate regime, $0.4<J_{2}/J_{1} <0.7$, most of the low-lying eigenstates are singlet ($S_{tot}=0$) and a few are triplet 
($S_{tot}=1$). The magnetic specific heat and the $z$-component of magnetic susceptibility, calculated with lowest six eigenstates, for the  layered lattice
 $(4\times 3)\oplus(4\times 3)$ are presented in the following.
 
\subsubsection{Specific Heat}
It has earlier been observed
\cite{bacci_prb_91_3,bishop_prb_08_3} that in a $J_{1}-J_{2}$ FAFH Model, as the frustration parameter is increased over the interval $0<J_{2}/J_{1}\le 0.5$, the systme goes from  ordered N\'{e}el state to disordered paramagnetic state; correspondingly,  the peak of the specific heat curve sharpens and its position shifts to lower temperatures. With further increase in the frustraion parameter $J_{2}/J_{1}$ beyond $0.5$, the system goes from disordered paramagnetic state to ordered collinear state and correspondingly, the peak of the specific heat broadens and its position shifts to higher temperatures. Accordingly, the above distinct behaviour of the specific heat peak {\it vs} temperature, as the system goes from  disordered to ordered state or {\it vice-veras} can be extended to examine the role of $J_{\perp}/J_{1}$ on the inplane magnetic order. \\
\begin{figure}[]
\centering
\subfigure[Specific heat {\it vs} temperature for different values of $J_{\perp}/J_{1}$ at $J_{2}/J_{1}=0.2$.]{\includegraphics[width=0.45\textwidth]{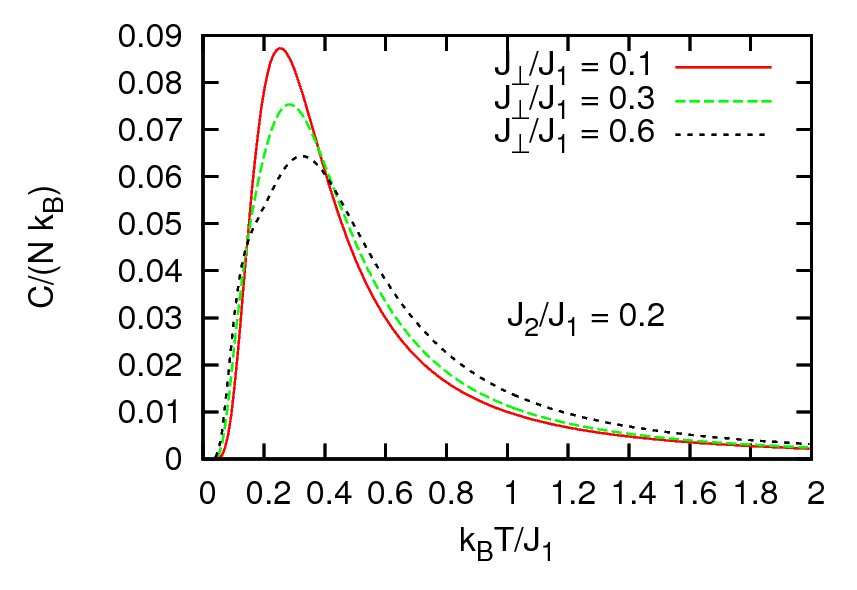}} \quad
\subfigure[Specific heat {\it vs} temperature for different values of $J_{\perp}/J_{1}$ at $J_{2}/J_{1}=0.6$. 
The position of specific heat peak changes little with respect to temperature 
as the interlayer coupling is increased from small to large values as $J_{\perp}/J_{1}=0.1,0.3,0.6$. The specific heat curve is sharpest for $J_{\perp}/J_{1}=0.6$, the highest of the values considered.]{\includegraphics[width=0.45\textwidth]{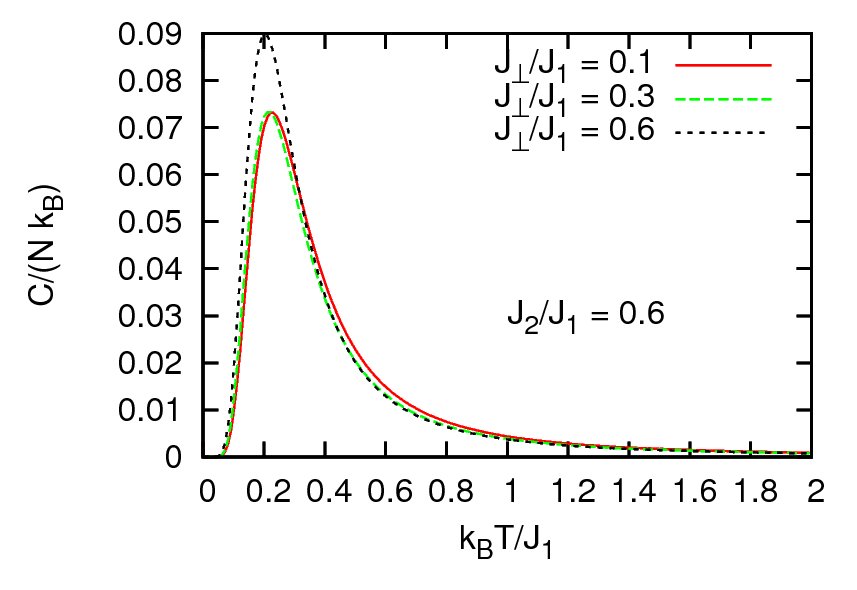}} 
\subfigure[Specific heat {\it vs} temperature for different values of 
$J_{\perp}/J_{1}$ at $J_{2}/J_{1}=0.8$. The peak position of specific heat 
changes little for small values of interlayer coupling $J_{\perp}/J_{1}=0.1,
0.3$. For higher values of $J_{\perp}/J_{1}=0.6$, the peak shifts to higher temperature. The specific heat curve is sharpest for $J_{\perp}/J_{1}=0.1$, the smallest value considered.]{\includegraphics[width=0.45\textwidth]{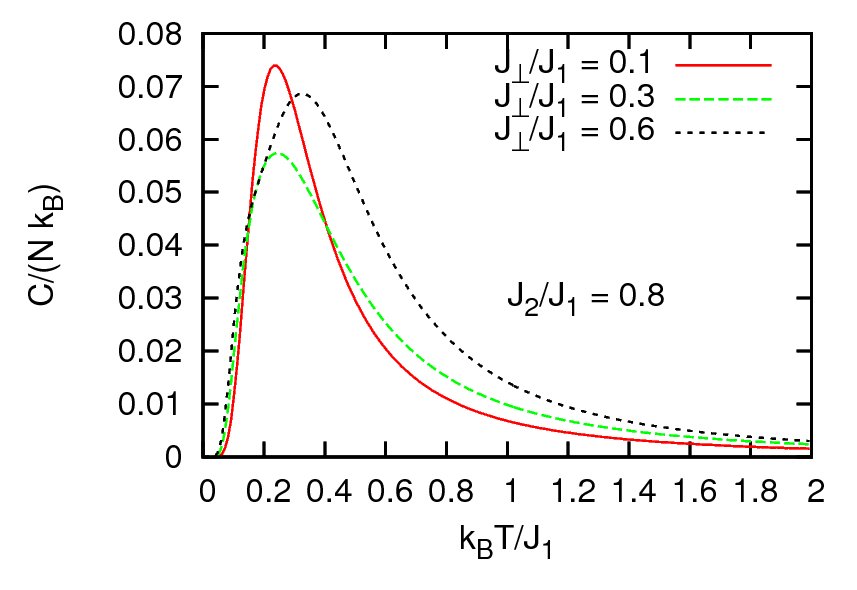}}
\caption{The dimensionless specific heat per spin $\left(C/\left(Nk_{B}\right)\right)$ {\it vs} scaled temperature $\left(k_{B}T/J_{1}\right)$ plots for $(4\times 3)\oplus(4\times 3)$ layered spin system for different values of $J_{\perp}$ at fixed values of inplane second-neighbor coupling $J_{2}/J_{1}$.}
\label{24spn_sh_zeta}
\end{figure}
\begin{figure}[]
\centering
\subfigure[Specific heat {\it vs} temperature for different values of $J_{2}/J_{1}$ at $J_{\perp}/J_{1}=0.1$. The specific heat peaks are sharpest at $J_{2}/J_{1}=0.4$ and $0.7$, the boundaries  of quantum paramagnetic region.]
{\includegraphics[width=0.45\textwidth]{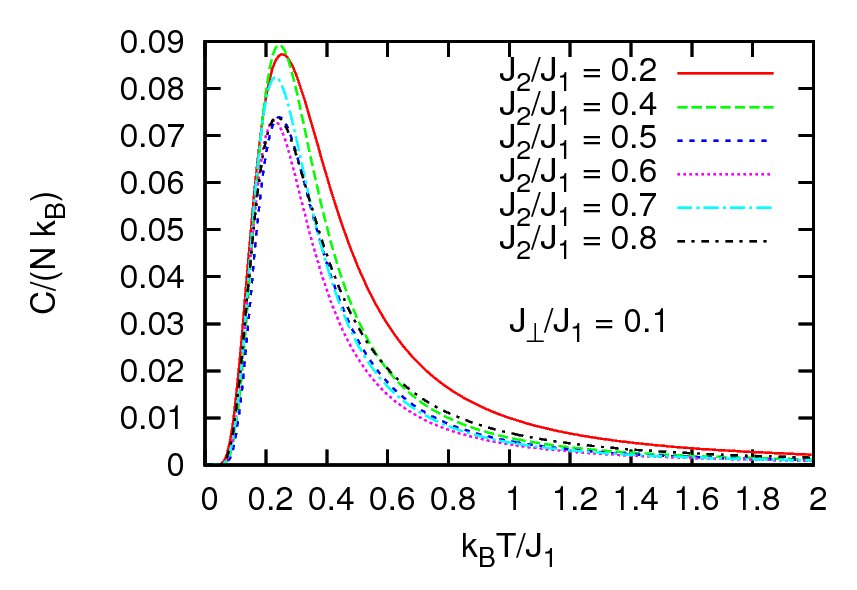}} \qquad
\subfigure[Specific heat {\it vs} temperature for different values of $J_{2}/J_{1}$ at $J_{\perp}/J_{1}=0.3$. 
 The specific heat peak is sharpest at $J_{2}/J_{1}=0.5$.]
 {\includegraphics[width=0.45\textwidth]{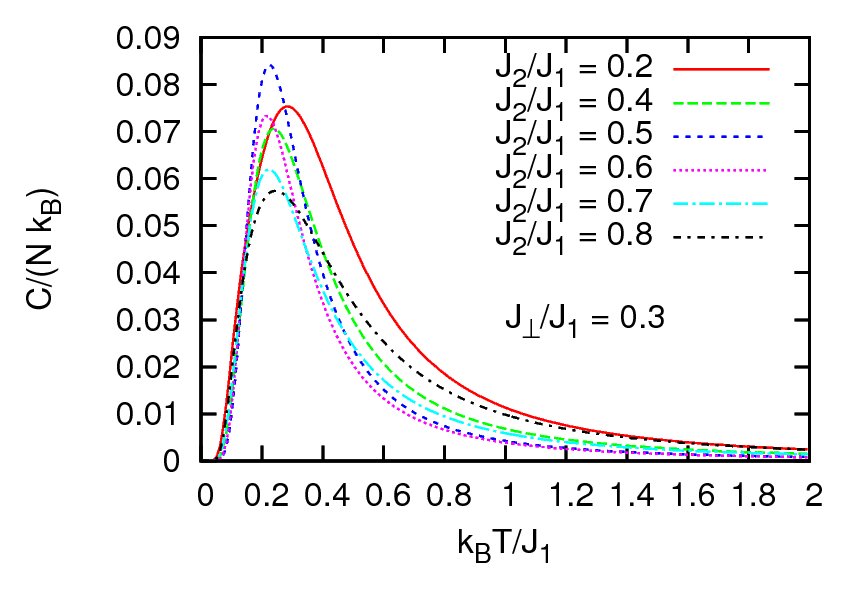}} 
\subfigure[Specific heat {\it vs} temperature for different values of $J_{2}/J_{1}$ at $J_{\perp}/J_{1}=0.6$. 
The specific heat peak is sharpest at $J_{2}/J_{1}=0.5$ and $0.6$. 
The boundaries of quantum paramagnetic region seem to have narrowed to $0.5\leq J_{2}/J_{1}\leq 0.6$.]
{\includegraphics[width=0.45\textwidth]{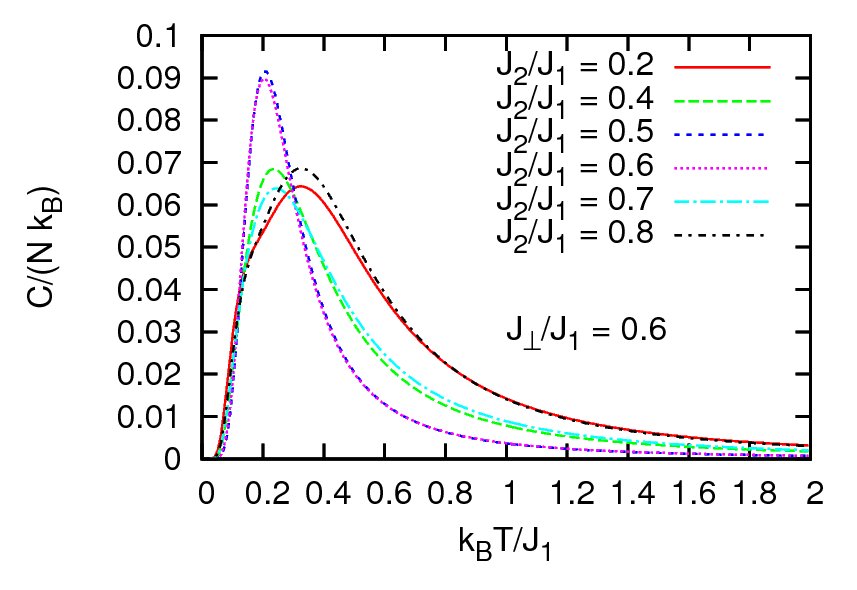}} 
\end{figure}
\begin{figure}[] 
\subfigure[Specific heat {\it vs} temperature for different values of $J_{2}/J_{1}$ for a $4\times 3$ lattice of $12$ spins. 
The specific curve is sharp for $J_{2}/J_{1}=0.4,\ 0.6$ but sharpest for $J_{2}/J_{1}=0.5$. 
The system is known to have quantum paramagnetically disordered for $0.4\leq J_{2}/J_{1}\leq 0.6$.]
{\includegraphics[width=0.45\textwidth]{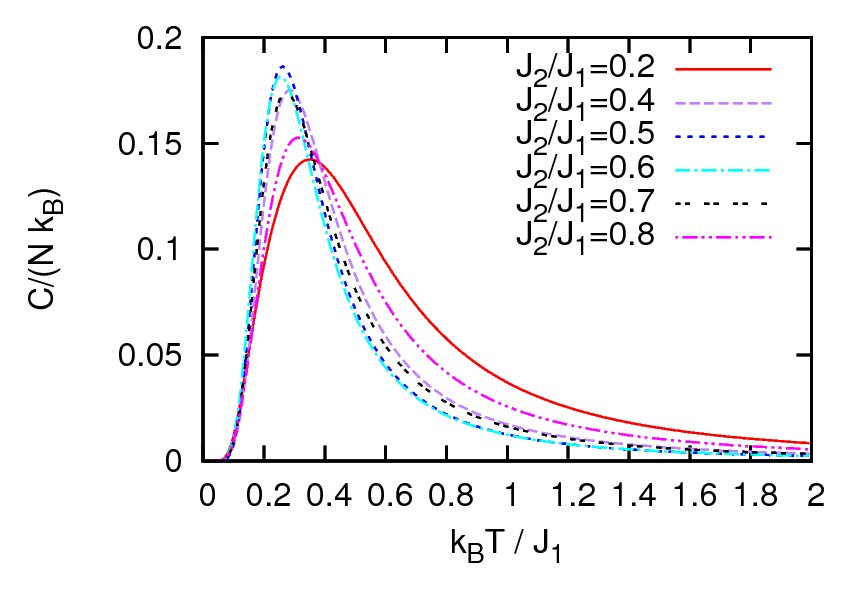}}
\caption{The dimensionless specific heat per spin $\left(C/\left(Nk_{B}\right)\right)$ {\it vs}  scaled temperature $\left(k_{B}T/J_{1}\right)$ plot of $\left((4\times 3)\oplus(4\times 3)\right)$ layered spin lattice for different values of inplane second-neighbor coupling $J_{2}/J_{1}$ at fixed values of $J_{\perp}/J_{1}$. 
For increasing values of $J_{2}/J_{1}$, the specific heat peak first shifts to lower temperatures and then (reversing the trend) to higher temperatures.}
\label{24spn_sh_alfa}
\end{figure}
In Figure(\ref{24spn_sh_zeta}a), we present specific-heat {\it vs} temperature plot 
with inplane frustration parameter $J_{2}=0.2J_{1}$ lying in the regime of semi-classical N\'{e}el ordered state in a pure 2D 
antiferromagnetic $J_{1}-J_{2}$ model \cite{richter_epjb_10_3,reuther_prb_11_3,bishop_prb_88}. The specific-heat peak monotonically shifts to higher 
temperatures as the interlayer coupling is increased from small to large values  {\it i.e.} $J_{\perp}/J_{1}=0.1,0.3,0.6$. The specific heat curve is sharpest 
for $J_{\perp}/J_{1}=0.1$, the smallest of the values considered. The specific-heat {\it vs} temperature curve flattens and the height of the peak 
decreases with increase in interlayer coupling. 
From calculations on small lattice of spins such as the one in the present work, it is difficult to 
 conclude whether the peak of the specific-heat corresponds to any phase  
transition or is just an energy crossover. Similarly, in Figure(\ref{24spn_sh_zeta}c) with inplane frustration parameter $J_{2}=0.8J_{1}$ lying in the regime of semi-classical collinear ordered state for a pure 2D antiferromagnetic $J_{1}-J_{2}$ model \cite{richter_epjb_10_3,reuther_prb_11_3,bishop_prb_88}, the 
specific-heat {\it vs} temperature curve again flattens and the position of the peaks shifts to higher temperatures with increase in $J_{\perp}$.\\
\indent
However, in Figure(\ref{24spn_sh_zeta}b) with inplane frustration parameter $J_{2}=0.6 J_{1}$ lying in the regime of quantum paramagnetic disordered  state for a pure 2D antiferromagnetic $J_{1}-J_{2}$ model \cite{richter_epjb_10_3,reuther_prb_11_3,bishop_prb_88}, the specific-heat {\it vs} temperature shows an entirely different behavior: no shifting of peak position is observed when the interlayer coupling $J_{\perp}$ is increased from $0.1J_{1}$ to $ 0.3J_{1}$ and $0.6J_{1}$.\\
\indent 
Thus as the interlayer coupling is increased, the system is driven to an  ordered state namely the N\'{e}el oreder state or the collinear ordered state, depending on the value of the inplane frustration parameter $J_{2}/J_{1}$. On the other hand, when the system is already in paramagnetic disordered state, an increase in interlayer coupling leads to no chnange in inplane magnetic order.  \\ 
\indent
In figure(\ref{24spn_sh_alfa}), we present magnetic specific heat {\it vs} temperature at given values of $J_{\perp}/J_{1}$ for different values of inplane frustration parameter. 
We observe in Figure(\ref{24spn_sh_alfa}a) that for $J_{\perp}=0.1J_{1}$, the specific heat peaks are sharp and occur at $k_{B}T/J_{1}=0.222 $ in the quantum paramegnetic regime corresponding to inplane frustration parameter values $J_{2}/J_{1}=0.4,0.5,0.6,0.7$; the peak occures at lower temperatures compared to other values of $J_{2}/J_{1}=0.2, 0.8$ lying in the N\'{e}el ordered and collinear ordered state, respectively. 
At $J_{\perp}=0.3J_{1}$ in Figure(\ref{24spn_sh_alfa}b), the specific heat peaks for $J_{2}/J_{1}=0.5,\ 0.6,\ 0.7$ corresponding to paramagnetic regime are sharp and occur at $k_{B}T/J_{1}=0.212 $, again lower than the temperatures at which specific heat curve peaks for $J_{2}/J_{1}=0.2,0.4$ corresponding to N\'{e}el ordered state and $J_{2}/J_{1}=,0.8$ corresponding to collinear ordered state.  
At significantly higher value of interlayer coupling $J_{\perp}/J_{1}=0.6$, we observe in Figure(\ref{24spn_sh_alfa}c) that the shepecific heat peaks flaten and occur at higher temperatures except for $J_{2}/J_{1}=0.5,\ 0.6$ corresponding to paramagnetic regime, for which the peaks are sharper and occur at a lower temperature $k_{B}T/J_{1}=0.212$ implying the shrinkig of quantum paramagnetic rgime with increase in interlayer coupling.\\
\indent
For reference, we plot in Figure(\ref{24spn_sh_alfa}d) the magnetic specific heat {\it vs} temperature for different values of inplane frustration parameter $J_{2}/J_{1}$ for a $4\times 3$ lattice of 12 spins. As $J_{2}/J_{1}$ is increased from $0.2$ to $0.4,\ 0.5,\ 0.6 $, the specific heat peaks sharpen and shift to lower temperatures. However, as it is further increased to $J_{2}/J_{1}=0.7,0.8$, the specific heat peak begins to flaten and shift to higher temperatures.\\
\indent
Thus the above observations on specific heat {\it vs} temperature curve can be summarized as follows: at small value of interlayer coupling $J_{\perp}/J_{1}=0.1$, there is a quantum paramagnetic order in $0.4<J_{2}/J_{1}<0.7$ regime whereas at $J_{\perp}/J_{1}=0.6$, the quantum paramagnetic regime narrows to $0.5<J_{2}/J_{1}<0.6$, implying the tendency of the paramagnetic regime to vanish with increasing interlaeyr coupling. For the finite lattice like the one in the present study, we do not expect to observe the vanishing of quantum paramagnetic regime as the interlayer coupling is incresaed due to finite size effects.
\subsubsection{Magnetic Susceptibility}
\begin{figure}
\centering
\subfigure[Magnetic Susceptibility {\it vs} temperature for different values of $J_{2}/J_{1}$ at $J_{\perp}/J_{1}=0.1$.]{\includegraphics[width=0.45\textwidth]{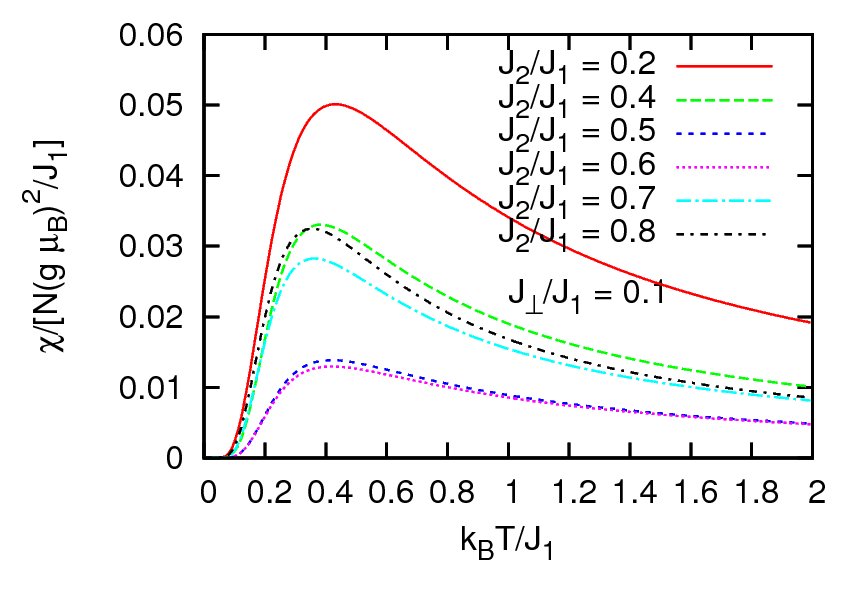}}
\label{24spn_chi_zta_pt1}
\subfigure[Magnetic Susceptibility {\it vs} temperature for different values of $J_{2}/J_{1}$ at $J_{\perp}/J_{1}=0.3$.]{\includegraphics[width=0.45\textwidth]{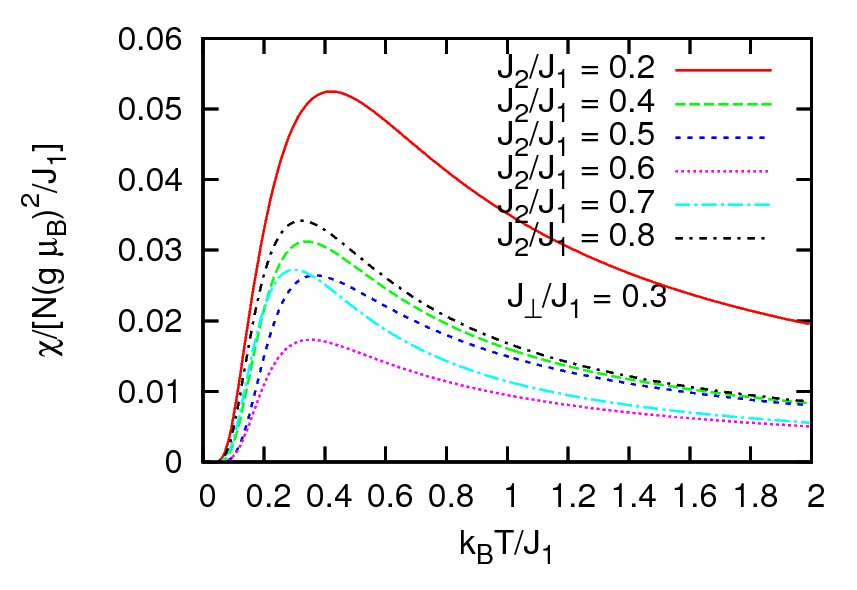}}
\label{24spn_chi_zta_pt3}
\subfigure[Magnetic Susceptibility {\it vs } temperature for different values 
of $J_{2}/J_{1}$ at $J_{\perp}/J_{1}=0.6$.]{\includegraphics[width=0.45\textwidth]{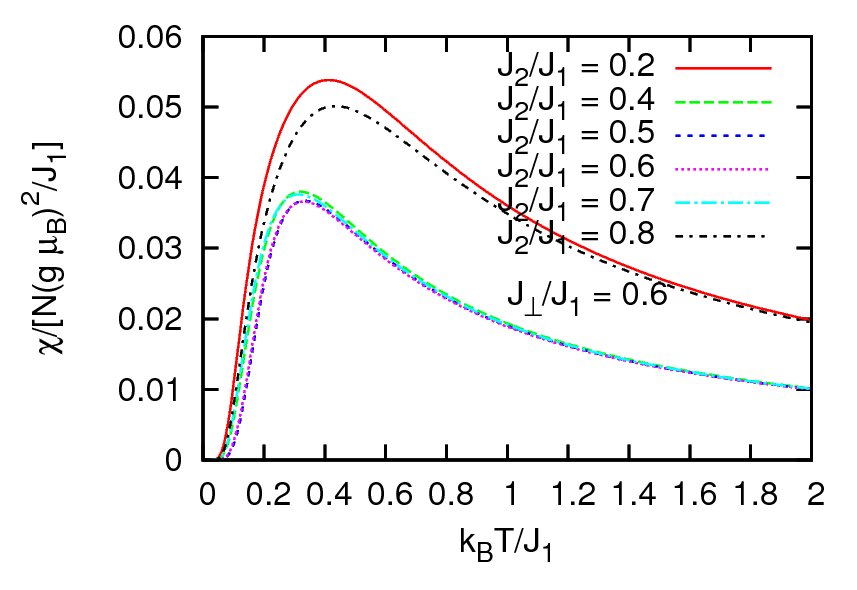}}
\caption{The dimensionless Magnetic Susceptibility per spin 
$ \chi /\left[N\left(g\mu_{B}\right)^{2}/J_{1} \right]$ {\it vs} scaled 
temperature $k_{B}T/J_{1}$ plot for $(4\times 3)\oplus (4\times 3)$ 
layered spin system for different values of inplane second neighbor coupling
 $J_{2}$ at three values of interlayer coupling $J_{\perp}$. For a given 
$J_{\perp}$, the magnetic susceptibility decreases with increasing $J_{2}$, 
reverses the trend and then increases with further increase in $J_{2}$, in the
entire range of temperature considered in the figure.}
\label{24spn_suscept_zta}
\end{figure}
Figure(\ref{24spn_suscept_zta}) presents the z-component of magnetic susceptibility {\it
vs} temperature for different values of inplane frustration parameter 
$J_{2}/J_{1}$ at three values of interlayer coupling $J_{\perp}/J_{1}$. 
In Figure(\ref{24spn_suscept_zta}a), we observe that at small value of interlayer coupling $J_{\perp}/J_{1} = 0.1 $, the system tends to shift towards magnetically disordered state leading to quantum paramagnetic order and the magnetic susceptibility reduces with increase in frustration parameter from small values $J_{2}/J_{1}=0.2$ corresponding to the N\'{e}el ordered state to intermediate values $J_{2}/J_{1}=0.4,\ 0.5,\ 0.6$ corresponding to paramagnetically disordered state. The magnetic susceptibilities corresponding to $J_{2}/J_{1}=0.5 $ and $0.6,$ almost coincide and remain lowest in the entire range of temperature considered in the figure.\\
\indent  
At a higher value of interlayer coupling $J_{\perp}/J_{1} = 0.3 $ in Figure(\ref{24spn_suscept_zta}b), the magnetic susceptibility 
for $J_{2}/J_{1}= 0.2$ and $J_{2}/J_{1}= 0.8 $ corresponding to the N\'{e}el and collinear ordered state respectively increases slightly but the  susceptibility of the paramagneticaly disordered state corresponding to $J_{2}/J_{1}= 0.4 \ 0.5$ and $0.6$ is enhanced significantly; the magnetic susceptibility curve corresponding to $J_{2}/J_{1}=0.6$ remains lowest in comparision to susceptibility curves for other values of $J_{2}/J_{1}$, in the entire range of temperature considered in the figure.\\
\indent
With further increase in interlayer coupling with $J_{\perp}/J_{1}= 0.6 $ in Fig.(\ref{24spn_suscept_zta}c), the peak value of magnetic susceptibility curves for the paramagnetically disordered state corresponding to $J_{2}/J_{1}= 0.4$, $J_{2}/J_{1}= 0.5$ and $J_{2}/J_{1}= 0.6$ is further enhanced and almost coincides with each other. Furthermore, the magnetic susceptibility curve is enhanced significantly for $J_{2}/J_{1}=0.8$ corresponding to collinear ordered phase. However, the magnetic susceptibility corresponding to $J_{2}/J_{1}=0.2$ remains large and of the same order of magnitude regardless of the values of $J_{\perp}/J_{1}$ in the entire range of temperatures considered in the figures(\ref{24spn_suscept_zta} a,b,c)\\
\indent
If we look at the position of the magnetic susceptibility humps for different values of inplane frustration parameter $J_{2}/J_{1}$, it shifts towards lower temperatures with increase in $J_{2}/J_{1}$ exhibiting signature of quantum paramagnetic disorder, as lesser thermal energy is required to cause thermal disorder. After further increase in $J_{2}/J_{1}$, the system re-orders and the hump begins to shift towards higher temperatures.\\
\begin{figure}[h!]
\centering
\subfigure[ Magnetic Susceptibility {\it vs} temperature for different values of $J_{\perp}/J_{1}$ at $J_{2}/J_{1}=0.2$.]{\includegraphics[width=0.45\textwidth]{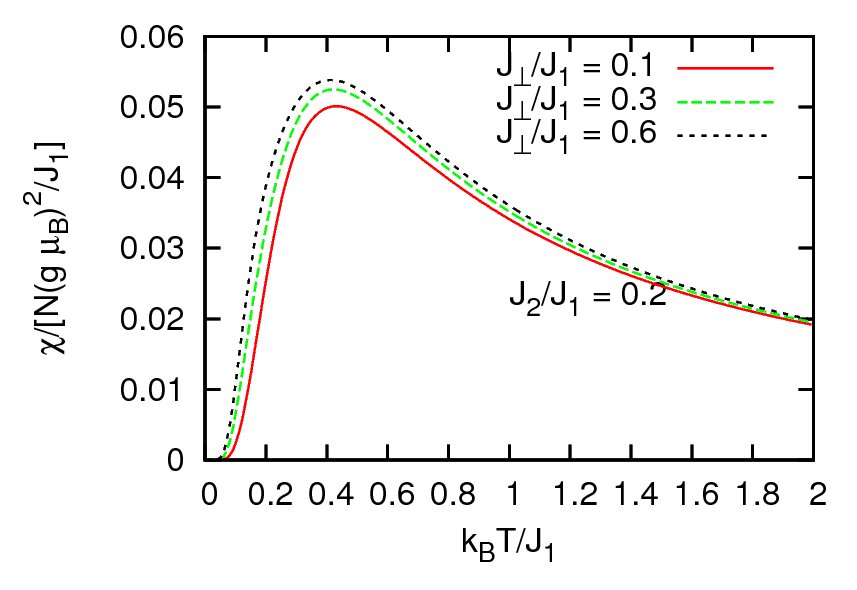}}
\label{24spn_chi_alf_pt2}
\subfigure[ Magnetic Susceptibility {\it vs} temperature for different values of $J_{\perp}/J_{1}$ at $J_{2}/J_{1}=0.6$.]{\includegraphics[width=0.45\textwidth]{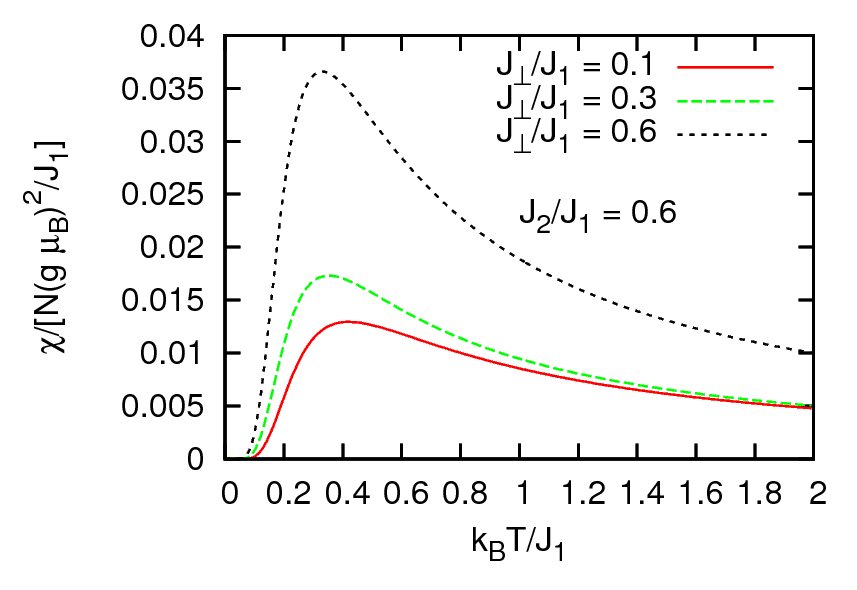}}
\label{24spn_chi_alf_pt6}
\subfigure[ Magnetic Susceptibility {\it vs} temperature for different values of $J_{\perp}/J_{1}$ at $J_{2}/J_{1}=0.8$.]{\includegraphics[width=0.45\textwidth]{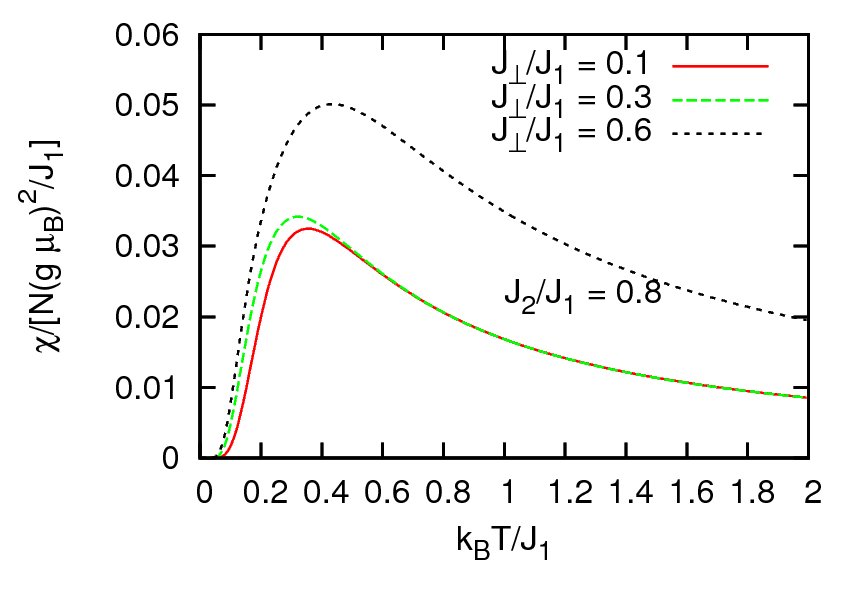}}
\label{24spn_chi_alf_pt8}
\caption{The dimensionless Magnetic Susceptibility per spin 
$ \chi /\left[N\left(g\mu_{B}\right)^{2}/J_{1} \right]$ {\it vs} scaled 
temperature $k_{B}T/J_{1}$ plot for $(4\times 3)\oplus (4\times 3)$ 
layered spin system for different values of interlayer coupling $J_{\perp}$ at
three values of inplane second neighbor coupling  $J_{2}$.}
\label{24spn_suscept_alfa}
\end{figure}
In Fig.(\ref{24spn_suscept_alfa}), we present magnetic susceptibility {\it vs} temperature for different values of interlayer coupling $J_{\perp}/J_{1}$ at three given values of inplane frustration parameter $J_{2}/J_{1}$.\\
\indent 
In Fig.(\ref{24spn_suscept_alfa}a) with 
$J_{2}/J_{1}=0.2$, the system is believed to exist predominantly in $S(\pi,\pi,\pi)$ ordered state corresponding to semiclassical N\'{e}el order and consequently we do not observe significant variance in magnetic susceptibility curve with increase in interlayer coupling $J_{\perp}/J_{1}$, in the entire range of temperature considered in the figure.\\
\indent
In Fig.(\ref{24spn_suscept_alfa}b) with $J_{2}/J_{1}=0.6$, the system is believed to exist in quantum paramagnetic disordered state and hence the magnetic susceptibility for small values of interlayer coupling $J_{\perp}/J_{1}=0.1, 0.3$, takes small values in the entire range of temperature considered in the
figure. When the interlayer coupling is increased to large value of $J_{\perp}/J_{1}=0.6$, the system re-orders ({\it i.e.}
the quantum paramagnetic disorder disappears and the 
magnetic susceptibility increases significantly in the
entire range of temperatures considered in the figure.\\
\indent
In Fig.(\ref{24spn_suscept_alfa}c) with $J_{2}/J_{1}=0.8$ the system is believed to exist in $S(\pi,0,\pi)$ ordered state, corresponding to semi-classical collinear order and the magnetic susceptibility is observed to be higher even at small values of interlayer coupling $J_{\perp}/J_{1}=0.1,0.3$ with respect to their values with 
$J_{2}=0.6J$ in Fig.(\ref{24spn_suscept_alfa}b). However, when the
interlayer coupling is increased to large value of $J_{\perp}/J_{1}=0.6$, the magnetic susceptibility increases significantly in the entire range of temperatures considered in the figure.\\
\indent 
The magnetic susceptibility corresponding to $J_{\perp}/J_{1}=0.6$ remains large and of the same order of magnitude regardless of the values of $J_{2}$ in the entire range of temperatures considered in the figures(\ref{24spn_suscept_alfa} a,b,c). If we observe the magnetic susceptibility hump, it moves towards higher temperatures implying that the system is being driven to a more ordered state as we increase the interlayer coupling, in agreement with Mermin-Wagner theorem\cite{mermin_prl_66_3}.\\ 
\indent 
The increment in magnetic susceptibilities of paramagnetically disordered state with increase in $J_{\perp}$ implies that the quantum parmagnetic ordered state tends to vanish and the spins re-order themselves as the interlayer coupling is increased. 
\section{Summary of Results and Discussion }  
We have presented an exact diagonalization study on $J_{1}-J_{2}-J_{\perp}$ model for a $24$-spin layered lattice in $\left(4\times 3\right)\oplus\left(4\times 3\right)$ geometry to examine the role of interlayer coupling on the inplane magnetic order.
For a quasi-2D $J_{1}-J_{2}$ model, the paramagnetic region is found to extend from $J_{2}=0.4J_{1}$ to $J_{2}=0.7J_{1}$.
 As we increase the interlayer coupling, the value of spin-gap in the paramagnetic region reduces since the interlayer coupling stabilizes the inplane order and thereby lowers the degree of disorder in paramagnetic regime. 
 We further observe that the spin-gap {\it vs} $J_{2}/J_{1}$ curves coincide for $J_{\perp}/J_{1}=0.6$ and $0.7$, implying that, $J_{\perp}=0.6J_{1}$ is the saturation value of the interlayer coupling beyond which the spin gap does not change with $J_{\perp}/J_{1}$. For an multilayered lattice \cite{richter_prl_06_3}, the corresponding saturation value will be $J_{\perp}=0.3J_{1}$, half the value for our two layered lattice, after correcting for double counting.\\
\indent 
At zero temperature, for small values of interlayer coupling, the inplane first-neighbor SSC becomes maximum in paramagnetic region {\it i.e.} short range order is dominant in paramagnetic regime. 
As the interlayer coupling is increased, the inplane first-neighbor SSC reduces {\it i.e.} short range correlation decreases in quantum paramegnetic regime. 
This is corroborated by our results on static spin structure factor also. 
In the quantum paramagnetic regime, $S(\pi,\pi,\pi)$ corresponding to N\'{e}el order decreases and $S(\pi,0,\pi)$ corresponding to collinear order increases  slowly with $J_{2}/J_{1}$ when $J_{\perp}/J_{1}$ is small. 
However, when $J_{\perp}/J_{1}$ is increased to $0.6$, the change in $S(\pi,\pi,\pi)$ and $S(\pi,0,\pi)$ becomes steeper indicating that in the thermodynamic limit, $S(\pi,\pi,\pi)$ may crossover directly to $S(\pi,0,\pi)$ and the intervening quantum paramagnetic regime disappears.\\
\indent
We observe that the specific heat {\it vs} temperature curve is flat in ordered states like N\'{e}el or collinear state but as the system enters quantum paramagnetically disordered state, the specific heat {\it vs} temperature curve acquires a sharp peak. For small values of interlayer coupling, we observe a sharp peak in specific heat {\it vs} temperature curve for the inplane frustration parameter interval $0.4 \le J_{2}/J_{1} 
\le 0.7 $ corresponding to paramagnetically disordered state. However, as the interlayer coupling is increased to a significant value, say $J_{\perp}/J_{1}=0.6$, a sharp peak in specific heat {\it vs} temperature curve is observed in the narrowed interval $0.5\le J_{2}/J_{1}\le 0.6$, 
indicating the shrinking of quantum paramagmetically disordered regime on $J_{2}/J_{1}$ scale.\\
\indent
From the (z-component of) magnetic susceptibility {\it vs} temperature plot we observe that the value of magnetic susceptibility in ordered states like N\'{e}el or collinear state is higher compared to its value in quantum paramagnetically disordered state. In the paramagnetically disordered state corresponding to the inplane frustration parameter interval $0.4\le J_{2}/J_{1}\le 0.7$ for small values of interlayer coupling $J_{\perp}/J_{1}$, the peak height of magnetic susceptibility curve is small. As $J_{\perp}/J_{1}$ is increased, the height of the magnetic susceptibility peak increases in the above interval of inplane frustration parameter, implying that the interlayer coupling drives the system to an ordered state.\\   
\indent 
In summary, the results presented in preceding sections show that for small values of interlayer coupling $J_{\perp}/J_{1}$, the system is in (a) semi-classical N\'{e}el ordered state for $J_{2}/J_{1}<0.4$ (b) semi-classical collinear ordered state for $J_{2}/J_{1}>0.7$ and (c) quantum paramagnetically disordered state for $0.4 <J_{2}/J_{1} < 0.7$. As $J_{\perp}/J_{1}$ is increased \cite{richter_prl_06_3}, the interval of  quantum paramagnetic disordered state on $J_{2}/J_{1}$ scale, narrows and long range order sets in. It has earlier been reported \cite{richter_prl_06_3} that for a multilayered lattice with $J_{\perp}=0.3J_{1}$, the paramagnetically disordered state vanishes and as $J_{2}/J_{1}$ is increased, the system goes directly from semi-classical N\'{e}el order to semi-classical collinear order. In our study on finite lattice, on increasing $J_{\perp}/J_{1}$, the intervening paramagnetically disordered regime narrows on $J_{2}/J_{1}$ scale but does not vanish due to finite size effect.\\
\noindent
{\bf Acknowledgement:} MMH and MAHA acknowledge financial support from University Grants Commission vide F. No. 41-891/2012(SR). Partial financial support 
from Martin-Luther University is also acknowledged where a part of the code 
for the present work was developed.

\end{document}